\pdfoutput=1
\RequirePackage{ifpdf}
\ifpdf 
\documentclass[pdftex]{sigma}
\else
\documentclass{sigma}
\fi

\usepackage{dsfont}
\usepackage{mathrsfs}

\usepackage{nicefrac}
\usepackage{tikz}
\usepackage{slashed}
\usepackage{upgreek}

\numberwithin{equation}{section}

\newcommand{\A}{\mathbb{A}}
\newcommand{\C}{\mathbb{C}}

\newcommand{\CP}{\mathbb{CP}}

\newcommand{\PT}{\mathbb{PT}}

\renewcommand{\P}{\mathbb{P}}

\newcommand{\T}{\mathbb{T}}
\newcommand{\Z}{\mathbb{Z}}
\newcommand{\p}{\partial}
\newcommand{\dbar}{\bar\partial}
\newcommand{\e}{\mathrm{e}}

\newcommand{\cA}{\mathcal{A}}
\newcommand{\cB}{\mathcal{B}}
\newcommand{\cC}{\mathcal{C}}

\newcommand{\cN}{\mathcal{N}}
\newcommand{\cO}{\mathcal{O}}
\newcommand{\cP}{\mathcal{P}}
\newcommand{\cT}{\mathcal{T}}

\newcommand{\cV}{\mathcal{V}}
\newcommand{\cI}{\mathcal{I}}

\newcommand{\cZ}{\mathcal{Z}}
\newcommand{\fZ}{\mathcal{Y}}
\newcommand{\fM}{\mathfrak{M}}

\newcommand{\cW}{\widetilde{\mathcal{Z}}}
\renewcommand{\P}{\mathbb{P}}
\newcommand{\SL}{\mathrm{SL}}

\newcommand{\tr}{\mathrm{tr}}

\newcommand{\be}{\begin{gather}\label}
\newcommand{\ee}{\end{gather}}
\newcommand{\bea}{\begin{eqnarray}\label}
\newcommand{\eea}{\end{eqnarray}}
\newcommand{\la}{\langle}
\newcommand{\ra}{\rangle}

\newcommand{\sA}{{\scalebox{0.6}{$A$}}}
\newcommand{\sB}{{\scalebox{0.6}{$B$}}}

\begin{document}
\allowdisplaybreaks

\newcommand{\arXivNumber}{2203.08087}

\renewcommand{\thefootnote}{}

\renewcommand{\PaperNumber}{045}

\FirstPageHeading

\ShortArticleName{From Twistor-Particle Models to Massive Amplitudes}

\ArticleName{From Twistor-Particle Models to Massive Amplitudes\footnote{This paper is a~contribution to the Special Issue on Twistors from Geometry to Physics in honour of Roger Penrose. The~full collection is available at \href{https://www.emis.de/journals/SIGMA/Penrose.html}{https://www.emis.de/journals/SIGMA/Penrose.html}}}

\Author{Giulia ALBONICO~$^{\rm a}$, Yvonne GEYER~$^{\rm b}$ and Lionel MASON~$^{\rm a}$}

\AuthorNameForHeading{G.~Albonico, Y.~Geyer and L.~Mason}

\Address{$^{\rm a)}$~The Mathematical Institute, University of Oxford, Oxford OX2 6GG, UK}
\EmailD{\href{mailto:giulia.albonico@maths.ox.ac.uk}{giulia.albonico@maths.ox.ac.uk}, \href{mailto:lmason@maths.ox.ac.uk}{lmason@maths.ox.ac.uk}}

\Address{$^{\rm b)}$~Department of Physics, Faculty of Science, Chulalongkorn University,\\
\hphantom{$^{\rm b)}$}~Thanon Phayathai, Pathumwan, Bangkok 10330, Thailand}
\EmailD{\href{mailto:yjgeyer@gmail.com}{yjgeyer@gmail.com}}

\ArticleDates{Received March 19, 2022, in final form June 08, 2022; Published online June 19, 2022}

\Abstract{In his twistor-particle programme of the 1970's, Roger Penrose introduced a representation of the massive particle phase space in terms of a pair of twistors subject to an internal symmetry group. Here we use this representation to introduce a chiral string whose target is a complexification of this space, extended so as to incorporate supersymmetry. We~show that the gauge anomalies associated to the internal symmetry group vanish only for maximal supersymmetry, and that correlators in these string models describe amplitudes involving massive particles with manifest supersymmetry. The models and amplitude formulae exhibit a double copy structure from gauge theory on the Coulomb branch to gravity, although the graviton remains massless. The formulae are closely related to those obtained earlier by the authors expressed in terms of the polarised scattering equations.}


\Classification{81U20; 83C60; 32L25; 81T30; 81T13}

\begin{flushright}
\begin{minipage}{80mm}
\it Dedicated to Roger Penrose on the occasion of his 90th birthday and the recent award of his Nobel prize in physics, who has been a constant source of inspiration to us.
\end{minipage}
\end{flushright}

\renewcommand{\thefootnote}{\arabic{footnote}}
\setcounter{footnote}{0}

\section{Introduction}

The twistor-strings of Witten \cite{Witten:2003nn}, Berkovits \cite{Berkovits:2004hg} and Skinner \cite{Skinner:2013xp} led to remarkably compact amplitude formulae for tree amplitudes for both super Yang--Mills \cite{Roiban:2004yf} and supergravity theories \cite{Cachazo:2013iaa, Cachazo:2012pz, Cachazo:2012kg,Geyer_2014}. These twistor-strings theories and formulae, including the worldsheet formulae of Cachazo, He and Yuan \cite{Cachazo:2013hca, Cachazo:2013iea, Cachazo:2014xea},
can all be understood under the umbrella of ambitwistor-strings \cite{Mason:2013sva}; a family of quantum field theories of holomorphic maps from a Riemann surface to the complexification of phase spaces of massless particles. These can come in various different twistorial or non-twistorial representations, with and without spin or supersymmetry and in different numbers of space-time dimensions. They exhibit a number of beautiful and perhaps unexpected features, such as a uniform simple structure in terms of residues in the moduli space~$\fM_{0,n}$ of $n$-points on $\CP^1$ modulo M\"{o}bius transformations, and one of the most direct realizations \cite{Bjerrum-Bohr:2016axv,Cachazo:2013iea, Monteiro:2013rya, Naculich:2014naa} of the double-copy between gauge and gravity amplitudes \cite{Bern:2008qj,Bern:2010ue}.

In their original form, all these ambitwistor-string theories and associated worldsheet formulae appear to be tightly restricted to theories and amplitudes involving only massless particles. However the underlying approach suggests that,
if one wishes to compute amplitudes for theories with massive particles, one should consider quantum field theories of holomorphic strings whose target is the complexified phase space of massive particles. The massless case has also shown that, in order to incorporate fermions simply, we should use a twistor representations of the phase space.

Nearly 50 years ago, Roger Penrose, followed by Zoltan Perjes, gave a twistor description of massive particles in terms of a set of two or more twistors up to an internal symmetry group \cite{Penrose:1974di,Perjes}. He proposed the \emph{twistor-particle programme} based on the twistor quantization of this description. In particular, it was hoped that the representation theory of the internal symmetry group should classify elementary particles; see for example \cite{Hughston:1979pg, Hughston_1981,TwistorProgramme, Perjes:1976sy} and references therein. Although this programme has not been pursued further by the twistor community, the framework was taken up by other authors in the particle physics community. For these authors, quantization via a worldline Lagrangian approach was used leading to studies of the spectrum of such twistor particle models, often incorporating supersymmetry. Two-twistor particle models include \cite{Bette:2004ip,Bette:2005br,deAzcarraga:2014hda}, see also
\cite{Deguchi:2015iuw,Deguchi:2017kvk} for more recent studies that have a good number of references to the evolution of the subject, including \cite{Fedoruk:2014vqa,Fedoruk:2003td}.\footnote{Note also a two-twistor model along the lines of an ambitwistor string \cite{Kunz:2020fwy,Kunz:2021dug}, but focussed on massless particles.}
 Such worldline actions are a stepping stone to ambitwistor-string formulations. These are holomorphic strings whose target is a complexified phase space; the action being built directly from the holomorphic symplectic potential on such a complexified phase space~\cite{Mason:2013sva}.

A particular advantage of the two-twistor representation of the phase space of massive particles is that it reduces to the nonlinear massive phase space via a symplectic quotient from the vector space of a pair of twistors. Such a symplectic quotient can be done via BRST in the quantum field theory, and all computations can be performed in a linear free-field quantum field theory on the Riemann surface.
However, a key lesson from the massless cases is that, even if one is only interested in bosonic Yang--Mills or gravity, fermionic symmetries are needed on the worldsheet and supersymmetries on space-time to obtain simple uniform formulae incorporating all relevant helicities. We will see that these supersymmetries can also be introduced in the massive case, leading to simple compact formulae for amplitudes in otherwise complicated, non-linear gauge and gravity theories.

In its simplest approach, massive particles were understood in terms of a pair of 4d twistors $\big(Z_a,\bar Z_a\big)\in \T\times \T$, $a=1,2$. Each twistor $Z\in \T$ has four complex components, that according to more recent (and less Penrosian) conventions are written as $Z=\big(\lambda_\alpha,\mu^{\dot\alpha}\big)$, i.e., as a pair of 2-component spinors; $\bar Z$ is the ${\rm SU}(2,2)$ complex conjugate of $Z$, defining a dual twistor by $\bar Z=\big(\bar\lambda_{\dot\alpha},\bar \mu ^\alpha\big)$. This description of massive particles was defined up to an internal symmetry group ${\rm SU}(2)\times \C$, where the ${\rm SU}(2)$ acts conformally invariantly on the $a$ index. It can be understood
in more conventional terms as the stabilizer of the massive momentum in the Lorentz group, the little group, see for example the massive spinor-helicity framework of \cite{Arkani-Hamed:2017jhn}.\footnote{More generally, $n$-twistor descriptions were considered with symmetry groups containing ${\rm SU}(n)$; in the twistor particle programme, particle multiplets were to be understood via the representation theory of such internal symmetry groups. The quantization of massive worldline models based on these descriptions has been studied by a number of authors, see \cite{Deguchi:2015iuw,Okano_2017} and references therein.} The factor of $\C$ in the symmetry group breaks conformal invariance and determines the particle masses.\looseness=1

Here we complexify twistors so that the complex conjugate twistor $\bar Z$ becomes a dual twistor~$\tilde Z$ independent of $\bar Z$ giving the pair $Y=\big(Z,\tilde Z\big)\in \T_\C:=\T\times \T^*$.
We can also think of such a~complexified twistor as a Dirac twistor $Y=\big(\lambda_A, \mu^A\big)$, given as a pair of 4-component Dirac spinors $\lambda_A=\big(\lambda_\alpha,\tilde\lambda_{\dot\alpha}\big)$ and $\mu^A=\big(\tilde\mu_{\alpha},\mu^{\dot\alpha}\big)$. We will also incorporate supersymmetry extending $Y\rightarrow \fZ=\big(\lambda_A, \mu^A, \eta^I\big)$ with $\mathcal{N}$ additional fermionic components $\eta^I$. This description gives a~natural inner product $\fZ\cdot \fZ:=Z\cdot \tilde{Z}+\eta_I\eta^I$ from the duality between $\T$ and $\T^*$ and skew form~$\Omega_{IJ}$.

Our 4d massive twistorial models are given by holomorphic maps from the Riemann surface~$\Sigma$ to the complexified two-twistor description of massive particles. They consist of a pair of complexified twistor fields $\fZ_a(\sigma)=\big(\cZ_a(\sigma),\cW_a(\sigma)\big)$, $a=1,2$ taking values in worldsheet spi\-nors~$\sqrt{\Omega^{1,0}_\Sigma}$ . To reduce to the twistor representation of the massive particle phase space, we also gauge the currents $\big(\fZ_a\cdot\fZ_b, \lambda^2, \tilde \lambda^2 \big)$ that generate the (complexified) internal symmetry group ${\rm SL}(2)\times \C\times \tilde \C$. Here $\lambda^2:=\det(\lambda)$ and its conjugate $\tilde\lambda^2$ determine the squared mass of the massive momentum $P_{\alpha\dot\alpha}=\lambda_\alpha^a \tilde\lambda_{\dot\alpha}^b\epsilon_{ab}$.
Thus we arrive at the model
\begin{gather*}
S_{4d} =\int_\Sigma \fZ^a\cdot\dbar\fZ_a+A_{ab}\fZ^a\cdot\fZ^b + A\big(\lambda^2-j^H\big)+ \tilde A \big(\tilde{\lambda}^2- j^{H}\big)+S_m.
\end{gather*}
Here $S_m$ is some theory dependent additional worldsheet matter that in particular can give rise to a current $j^H$ associated to some symmetry generator $H$.
The $A_{ab}=A_{(ab)}$ are gauge fields for the ${\rm SL}(2)$ little group, and $\big(A,\tilde{A}\big)$ gauge the $\C\times \tilde \C$ part of the internal symmetry group; they are also Lagrange multipliers relating the values of the particle masses
to their charges under~$H$.

Although this two-twistor massive model is a string whose target is the complexified two-twistor description of massive particles of~\cite{Perjes}, it can be identified with the dimensional reduction of the 6d and 5d ambitwistor strings in~\cite{Geyer:2020iwz}.
The contractions of the massive spinor helicity variables correspond to two components of the internal momentum when embedding the massive variables in a six dimensional massless momentum.
The two-twistor string produces correlators that localize on delta functions that fix the values of internal momenta in terms of charges under~$H$ and~$\tilde H$ for all particles involved. In addition to this, the correlators are further localized by delta functions imposing a polarised version of the scattering equations as in \cite{Albonico:2020mge, Geyer:2018xgb}.

The models above not only allow us to derive formulae involving any number of massive particles, but also give an alternative formulation of the massless models in \cite{Geyer_2014}. This is of particular importance as it presents a framework in which a massless field can be \emph{deformed} to go off-shell, which is a necessary prerequisite for defining a gluing operator in the four dimensional twistorial model and producing loop amplitudes.

{\bf Paper summary.} In the next section we introduce the two-twistor geometry of the massive particle phase space. We then briefly present the Penrose transform and its complexification. This can be used in a two-twistor string that computes amplitudes for theories with massive particles in four dimensions. Such formulae incorporate fermions and supersymmetry, generalizing the massless case of \cite{Geyer_2014}. They are based on the polarised scattering equations that have already been introduced and studied in six and five dimensions \cite{Albonico:2020mge,Geyer:2019ayz,Geyer:2020iwz}; these can also be related \cite{Schwarz:2019aat} to the formulae of \cite{Cachazo:2018hqa}.
We focus on a model adapted to the Coulomb branch of $\mathcal{N}=4$ SYM; this contains a gauge field, fermions and scalars that have a vacuum expectation value to give masses to some of the particles, analogous to the standard model. Nevertheless, as in the models of~\cite{Geyer:2015jch, Geyer:2020iwz}, we can write down a full range of models for particles of spin-0, spin-1 and spin-2 following the double copy, although the scope for introducing masses into the gravity models are limited.

\section{Massive particles}

We first review the twistor description of massive particles in terms of a pair of twistors with redundancy described by the two-twistor internal symmetry group ${\rm SU}(2)\times \C$. This framework ties in directly with the (more recent) spinor-helicity formalism for expressing polarization data for massive particles.
 Anticipating the string model we then complexify the two-twistor description, and introduce the Penrose transform for massive momentum eigenstates.

{\bf Conventions and notation.} We will work with spinors in the complex, so that the Lorentz group in $4$ dimensions is double-covered by ${\rm SL}(2,\C)\times {\rm SL}(2,\C)$ with each factor acting separately on the positive and negative chirality spinors. The positive and negative chirality spinors carry indices $\alpha=1,2$, $\dot\alpha=\dot 1, \dot 2$, raised and lowered with $\varepsilon_{\alpha\beta}=\varepsilon_{[\alpha\beta]}$, $\varepsilon_{12}=1$ and its inverse, as well as the corresponding dotted version. We denote spinor inner products by the conventional bracket notation,
\begin{gather*}
\la\lambda_1\lambda_2\ra=\lambda_{1\alpha}\lambda_2^\alpha, \qquad
[\lambda_1\lambda_2]=\Tilde{\lambda}_{1\dot \alpha}\Tilde{\lambda}_2^{\dot \alpha}, \qquad
 \Tilde{\lambda}_{\dot \alpha}=\Tilde{\lambda}^{\dot \beta}\varepsilon_{ \dot \beta\dot \alpha}, \qquad \Tilde{\lambda}^{\dot \alpha}=\varepsilon^{\dot \alpha \dot \beta}\Tilde{\lambda}_{\dot \beta}.
\end{gather*}

\subsection{Review of twistor internal symmetry groups for massive particles}

{\bf Massless particles.} As described in \cite{Penrose:1974di,Penrose:1986ca, Perjes, Perjes:1976sy}, a general twistor $Z^\cA=(\lambda_\alpha,\mu^{\dot\alpha})\in \T$
determines a massless particle whose momentum $P_{\alpha\dot\alpha}$ and angular momentum $M^{\mu\nu}=M^{\alpha\beta}\varepsilon^{\dot\alpha\dot\beta} + \text{c.c.}$ about the origin, can be assembled into the \emph{angular momentum twistor} given by
\begin{gather*}
L^{\cA\cB}:=\begin{pmatrix}
0&P_\alpha^{\dot\beta}\\P_\beta ^{\dot\alpha}& M^{\dot\alpha\dot\beta}
\end{pmatrix}=\begin{pmatrix}
0&\lambda_\alpha \bar\lambda^{\dot\beta}\\\lambda_\beta \bar \lambda^{\dot\alpha}& \bar\lambda^{(\dot\alpha}\mu^{\dot\beta)}
\end{pmatrix}=
Z^{(\cA} I^{\cB)\cC}\bar Z_\cC.
\end{gather*}
In this formula, the \emph{infinity twistor} breaks conformal invariance and is defined by
\begin{gather*}
I^{\cA\cB}\bar Z_\cB= \big(0,\bar \lambda^{\dot\alpha}\big), \qquad
I_{\cA\cB} Z^\cB= (0, \lambda^{\alpha}),
\end{gather*}
extending the spinor contractions to degenerate inner products $\langle Z_1 Z_2\rangle:=I_{\cA\cB} Z^\cA_1Z^\cB_2=\la\lambda_1\lambda_2\ra$ on twistor space. The angular momentum twistor is invariant under the internal symmetry transformation $Z\rightarrow \e^{{\rm i}\theta} Z $, which we can identify as the little group rotating the phase of the constituent spinors of the massless momentum $P_{\alpha\dot\alpha}=\lambda_\alpha\bar \lambda_{\dot\alpha}$.

{\bf Massive particles.} In order to describe massive particles, we introduce a sum over 
two twistors~$Z_a^\cA$, $a=1,2$ with complex conjugates $\bar Z_\cA^a$. These yield the angular momentum twistor
\begin{gather*}
L^{\cA\cB}=Z^{(\cA}_a I^{\cB)\cC}\bar Z_\cC^a.
\end{gather*}
 In particular, the momentum is given by
\begin{gather*}
P_{\alpha\dot\alpha}=\lambda_{a\alpha}\bar\lambda^a_{\dot\alpha},
\end{gather*}
and so we can identify the indices $a$, $b$ as the ${\rm SU}(2)$ little-group indices that stabilizes the massive momentum $P_{\alpha\dot\alpha}$ inside the Lorentz group.

Penrose~\cite{Penrose:1974di} and Perjes~\cite{Perjes} define the two-twistor internal symmetry group to be the Poincar\'e invariant transformations that preserve the angular momentum twistor. This group is ${\rm SU}(2)\times \C$, where the ${\rm SU}(2)$ acts as the massive little group, and the factor of $\C$ is given by the complex transformations
\begin{gather*}
\delta Z^\cA_a \propto I^{\cA\cB}\bar Z_\cB^b \epsilon_{ba}.
\end{gather*}
These symmetries all preserve the symplectic form and potential \cite{Tod:1977vf}
\begin{gather*}
\Omega_m:={\rm d}\Theta_m, \qquad
2\Theta_m:= {\rm i}Z^\cA_a {\rm d}\bar Z_\cA^a-{\rm i}\bar Z_\cA^a {\rm d}Z^\cA_a.
\end{gather*}
The internal symmetry group action with respect to the potential $\Theta_m$ is generated by the Hamiltonians
\begin{gather*}
Z^\cA_{(a}\cdot \bar Z_{b)\cA}, \qquad
\lambda^2:=\frac12 \lambda_\alpha^a\lambda_a^\alpha=\frac12
\la Z_a,Z^a\ra=\frac 12 I_{\cA\cB} Z^\cA_a Z^{\cB a},
\end{gather*}
 for the factors of ${\rm SU}(2)$ and $\C$ respectively. We can therefore define the 
 phase space $\cP_m$ for particles of mass $m$ as the symplectic quotient
 \begin{gather}\label{real-phase-space}
 \cP_{m}=\big\{ Z_a \in \T\times \T \mid Z_{(a}\cdot\bar Z_{b)}=0, \la Z_a Z^a\ra=m\big\}/ \{ {\rm SU}(2)\times \C\}.
 \end{gather}
It is easy to see that this is a 6 real-dimensional symplectic manifold with symplectic potential~$\Theta_m$.

\subsection{Dirac spinors and spinor-helicity for massive particles}

As remarked above, the ${\rm SU}(2)$ of the internal symmetry group is the massive particle ``little group'', the subgroup of the spin double cover of the Lorentz group that preserves a time-like momentum $P_{\alpha\dot\alpha}$; 
the representations of this little group are naturally identified with
 the polarization states of massive particles as follows. For
a massive particle of momentum $k_{\alpha\dot\alpha}$ we write as above
\begin{gather*}
k_{\alpha\dot\alpha}=\kappa_{a \alpha} \tilde{\kappa}_{\dot\alpha}^a,
\end{gather*}
where $a=1,2$ is an ${\rm SU}(2)$ little group index raised and lowered by $\varepsilon_{ab}=\varepsilon_{[ab]}$, $\varepsilon_{12}=1$.
In the real case $\tilde \kappa^a_{\dot\alpha}$ can be taken to be the complex conjugate of $\kappa_{\alpha a}$ reducing the little group to ${\rm SU}(2)$. We denote little group contractions by
\begin{gather*}
 (v_1v_2):=v_{1a}v_{2b}\varepsilon^{ab}.
\end{gather*}
 The mass $m$ is given by
$ k^2=m^2=\det (k_{\alpha\dot\alpha})= \det\kappa\det\tilde{\kappa}$;
so defining
\begin{gather*}
\det \kappa=M,\qquad
\det \tilde{\kappa}=\tilde M,
\end{gather*}
we have $M\tilde M=m^2$ and although we can fix the phases of $\kappa$ and $\tilde \kappa$ so that $M=\tilde M =\pm m$, later we will want to keep them independent before they are fixed by the model.

Massive particles are not chiral, and two-component spinors necessarily double up with their conjugates. For a more compact notation, we introduce Dirac 4-component spinors with indices denoted by capital Roman letters from the beginning of the alphabet as
\begin{gather*}
\psi_A=\big(\psi_\alpha,\tilde\psi^{\dot\alpha}\big), \qquad \psi^A=\varepsilon^{AB}\psi_A:=\big(\psi^\alpha,\tilde\psi_{\dot\alpha}\big), \qquad \psi_{1A}\psi_2^A= \psi_{1\alpha} \psi_2^\alpha+ \tilde \psi_{1}^{\dot\alpha}\tilde\psi_{2\dot\alpha},
\end{gather*}
and we will raise and lower indices with $\varepsilon^{AB}$, $\varepsilon_{AB}$, $\varepsilon^{AB}\varepsilon_{AC}=\delta^B_C$. Also note the $\gamma_5 $ matrix defined by
\begin{gather*}
\gamma_{5A}^B\psi_B={\rm i}\big(\psi_\alpha,-\tilde\psi^{\dot\alpha}\big).
\end{gather*}
The mass-$m$ Dirac operator $D^m_{AB}=D^m_{[AB]}$ in this notation is
\begin{gather*}
D^m_{AB}:=-{\rm i}\nabla_{AB}+m\varepsilon_{AB}=-{\rm i}\begin{pmatrix}
0&\nabla_{\alpha}^{\;\dot\beta}\\ -\nabla^{\dot\alpha}_{\;\beta}&0
\end{pmatrix}
+m\varepsilon_{AB}.
\end{gather*}
The spin $s$ massive field equations for $\Psi^{A_1\cdots A_2s}=\Psi^{(A_1\cdots A_2s)}$ becomes
\begin{gather*}
D^m_{BA_1}\Psi^{A_1\cdots A_2s}=0.
\end{gather*}
At spin $s=1$, we obtain $F_{AB}=F_{(AB)}$ whose $2\times 2$ block-decomposition contains the 2-form curvature spinors along the diagonal and $mA_{\alpha\dot\beta}$ on the off-diagonal, where $A_{\alpha\dot\alpha}$ is the one-form potential.

 Introducing a little group spinor $\epsilon_a$, the general plane wave on Minkowski space of spin-$s$ can be decomposed into Dirac spinor wave functions as
\begin{gather}\label{plane-wave}
\Psi_{A_1\cdots A_{2s}}=\epsilon_{A_1}\cdots \epsilon_{A_{2s}} \e^{{\rm i}k\cdot x}, \qquad \epsilon_A=\epsilon_a \kappa^a_A=:(\epsilon \kappa_A), \qquad \kappa_{Aa}=\big(\kappa_{\alpha a},\tilde\kappa^{\dot\alpha}_{a}\big).
\end{gather}
For spin $1/2$, this is an ordinary massive Dirac field momentum eigenstate with polarization~$\epsilon_a$; for spin $s=1$ this describes a massive field with potential $A_{\alpha\dot \alpha}= \frac1m \epsilon_{ab}\,\kappa^a_\alpha\tilde{\kappa}^b_{\dot \alpha} \,\e^{{\rm i}k\cdot x}$, with polarization $\epsilon_{(ab)}=\epsilon_a\epsilon_b$.
In general, spin-$s$ massive particles transform as the symmetric part of rank $2s$ tensors of the massive little group ${\rm SU}(2)$, with polarization data $\epsilon_{a_1\cdots a_{2s}}=\epsilon_{(a_1\cdots a_{2s})}$. Note that the polarization in \eqref{plane-wave} is taken to be simple to tie in with later supersymmetric expressions, corresponding to a null polarization vector.
We refer to \cite{Arkani-Hamed:2017jhn,Conde:2016izb,Conde:2016vxs} for more extended recent discussions of spinor-helicity for massive particles.

To reduce to the massless case, we can take half the spinor components to vanish $\kappa_{1\alpha}=0=\tilde \kappa_{0}^{\dot\alpha}$, whereupon the little group spinor components $\epsilon_0$ and $\epsilon_1$ parametrize the positive and negative helicity states respectively.

\subsection{The complexified particle phase space and Penrose transform}

In all (massless) ambitwistor strings \cite{Mason:2013sva}, the target space is the complexification of the massless particle phase space, often referred to as ambitwistor space and denoted by $\A$. To define this we first introduce
complexified twistor space $\T_\C$ by
\begin{gather*}
Y=\big(Z,\tilde Z\big)\in \T_\C:=\T\times \T^*.
\end{gather*}
Then the complexified phase space of massless particles in four dimensions has become known as ambitwistor space $\A$, defined non-projectively as the holomorphic symplectic quotient
\begin{gather}\label{A-def}
\A=\{ Y\in \T_\C\mid Y\cdot Y:=Z\cdot \tilde Z=0\}/ \big\{ Z\cdot\p_Z-\tilde Z\cdot \p_{\tilde Z}\big\},
\end{gather}
with respect to the symplectic structure
\begin{gather*}
\Omega_\A={\rm d} \Theta_\A, \qquad
\Theta_\A:= {\rm i}Z\cdot {\rm d}\tilde Z-{\rm i}\tilde Z\cdot {\rm d} Z.
\end{gather*}
This is the target of the original twistor strings \cite{Berkovits:2004hg, Skinner:2013xp, Witten:2003nn} and the closely related ambitwistor strings \cite{Geyer_2014}.

In analogy with the massless case, here we take the target space to be $\cP_m^\C$, the complexification of the massive particle phase space $\cP_m$.
We represent $\cP^\C_m$ as the holomorphic symplectic quotient analogue of \eqref{real-phase-space} as
\begin{gather*}
 \cP^\C_m:=\big\{Y_a\in \T^\C\times \T^\C\mid Z_{(a}\cdot \tilde Z_{b)}=0, \, \langle Z_a Z^a\rangle=\big[\tilde Z_a\tilde Z^a\big]=m\big\}/\SL(2,\C)\times \C\times \tilde \C.
\end{gather*}
One of the oldest applications of twistor theory has been to provide solutions to the free field equations. In the massless case this is achieved via the Penrose transform, which represents zero-rest mass helicity-$h$ fields as twistor cohomology classes $H^{1}(\P\T,\mathcal{O}(2h-2))$. Using the identifications $\A=T^*\PT=T^*\PT^*$, representatives of these cohomology classes can also be pulled back to ambitwistor space.
While two-twistor descriptions in the literature \cite{Hughston_1981} lead to~$H^2$ representatives by building on the real massive particle phase space, we use the complexification~$\cP^\C_m$ to obtain representatives in (Dolbeault) cohomology classes $H^1\big(\cP^\C_m, \cO(2s-2)\big)$ that couple naturally to the worldsheet. Here we will focus on the scalar case $s=0$, the extension to spinning particles can be achieved most straightforwardly via supersymmetry and is discussed in Section~\ref{sec:susy}.

To represent the plane wave \eqref{plane-wave} with momentum $k_{\alpha\dot\alpha}= \kappa_{\alpha a}\tilde{\kappa}_{\dot\alpha}^a$ on $\T^\C\times\T^\C$ it will be convenient to reorganise the spinor constituents of $Y_a$ as a ``Dirac twistor''
\begin{gather*}
Y_a=\big(\lambda_{aA},\mu^A_a\big), \qquad \lambda_{aA}:=\big(\lambda_{a\alpha},\tilde\lambda_{a}^{\dot\alpha}\big),\qquad \mu^A_a:=\big(\mu^{\dot\alpha}_a,\tilde\mu_{\alpha a}\big).
\end{gather*}
Writing $\kappa_{Aa}=\big(\kappa_{\alpha a},\tilde\kappa^{\dot\alpha}_{ a}\big)$, we define the corresponding cohomology representative in \linebreak $H^1\big(\cP^\C_m, \cO(-2)\big)$ by introducing four auxiliary complex variables $u_a$, $v_a$;
\begin{gather}\label{H1_plane-wave}
\Phi_{\kappa}(Y_a)=\int {\rm d}^2u\,{\rm d}^2v
 \bar\delta^4((u\lambda_A)-(v\,\kappa_A))\bar\delta((v,\epsilon)-1)
\exp \big(\big(u \mu^{A}\big)\epsilon_{A} \big).
\end{gather}
Here the line bundle $\cO(n)$ is the bundle of homogeneity degree $n$ in the $Y_a$, and for a complex variable $z$ we define $\bar\delta(z)$ to be the distributional $(0,1)$-form
\begin{gather*}
\bar\delta(z)= \dbar \frac{1}{2\pi {\rm i} z}= \delta(\operatorname{Re} z)\delta(\operatorname{Im} z)\, {\rm d}\bar z.
\end{gather*}
After the $u_a$, $v_a$ integrals have been performed, $\Phi_{\kappa}(Y_a)\in H^1\big(\cP^\C_m, \cO(-2)\big)$ is indeed a $(0,1)$-form as desired, and the integration over $u$ ensures invariance under the ${\rm SL}(2,\C)$ little group. For the Penrose transform, we take $\det(\kappa)$ and $\det(\tilde\kappa)$ to be unconstrained; in the two-twistor string these quantities will be constrained to agree with the particle masses determined by the underlying theory.

To see that this indeed corresponds to a plane-wave on space-time, we impose the incidence relations $\mu^{\dot\alpha}_a= {\rm i}x^{\alpha\dot\alpha}\lambda_{a\alpha}$ and $\tilde\mu^{\alpha}_a=- {\rm i}x^{\alpha\dot\alpha}\tilde\lambda_{a\dot\alpha}$. Then on the support of the delta functions we have $(u\lambda_A)=(v\kappa_A)$ and $(v\epsilon)=1$,
giving\footnote{This Penrose transform is closely related to the (indirect) 6d Penrose transform \cite{Mason:2011nw,S_mann_2013}. The twistor space for 6d is $\T_\C|_{Y\cdot Y=0}$ and the plane wave \eqref{plane-wave} is represented by
\begin{gather*}
\Psi_\kappa(Y)=\int (\epsilon v)^n\frac{{\rm d}s }{s^{n-1}} (v {\rm d}v) \bar\delta^4( s\lambda_A - (v\kappa_A))
\exp \big(s\mu^A\epsilon_A/(\epsilon v)\big) \in H^2( \P\A, \cO(n-2)).
\end{gather*}
Following \cite{Geyer:2020iwz, Mason:2012va, ReidEdwards:2012tq,S_mann_2014}, the massive $\Phi_{\kappa}(Y_a)$ in \eqref{H1_plane-wave} can then be constructed via $\Phi_{\kappa}(Y_a)= \int \Psi_\kappa((u,Y))(u{\rm d}u)$.}
\begin{gather}
\big(u \mu^{A}\big) \epsilon_{A}={\rm i} x^{\alpha\dot\alpha} (v\kappa_\alpha)(\epsilon\tilde \kappa_{\dot\alpha})-{\rm i}x^{\alpha\dot\alpha} (v\tilde \kappa_{\dot \alpha})(\epsilon\kappa_{\alpha}) ={\rm i} x\cdot k.\label{xdotk}
\end{gather}
The parameters $(u_a,v_a)$ can be integrated against the delta functions to yield the single delta function $\bar\delta\big(k\cdot P -m \lambda^2-m \tilde \lambda^2\big)$, where $P_{\alpha\dot\alpha}=\lambda_{a\alpha}\tilde\lambda_{\dot\alpha}^a$; this gives the Dolbeault representation for a simple pole. This delta function imposes a massive analogue of the \emph{scattering equations} that play a key role in the CHY formulae for massless amplitudes \cite{Cachazo:2013iaa, Cachazo:2012pz, Cachazo:2012kg,Geyer_2014}. A version of these \emph{massive scattering equations} has been studied for massive amplitudes in the CHY representation~\cite{Naculich:2015zha}, see the discussion in Section~\ref{sec:discussion} for more details. The delta-functions in $\Phi_\kappa$ have become known as the \emph{polarized scattering equations} due to their dependence on a choice of polarization spinor.\footnote{The polarization spinor will play a more prominent role in amplitude formulae based on the polarized scattering equations, because the path integrals introduce $\epsilon$-dependence in $\lambda_\sA(\sigma)$.} These serve to define additional and unique parameters $u_a$, $v_a$ on the support of the massive scattering equations that will play a key role later.

\subsection{Supersymmetric extension}
\label{sec:susy}
We will aim to generate supersymmetric formulae for two reasons. Amplitudes for theories with a variety of spins become drastically simpler in the supersymmetric case because many or all the particles can be expressed as one multiplet. This leads to uniform formulae from which different sectors with particles of different spins can be read off. Moreover, amplitudes for non-supersymmetric theories can be extracted from these superamplitudes at tree-level and at one-loop \cite{Geyer:2015jch,Johansson:2014zca}.
A more structural reason is that all (ambi-) twistor string models that describe gauge theory and gravity require space-time supersymmetry to be anomaly free -- the supersymmetric extension of twistor space includes additional fermionic variables that cancel anomalies from the bosonic variables. Thus we introduce a supersymmetric extension of $\mathcal{P}_m^{\mathbb{C}}$, as well as the plane wave $\Phi_\kappa$.

On the Coulomb branch of $\cN=4$ super Yang--Mills, some scalars acquire a vacuum expectation value, effectively breaking the gauge group from ${\rm SU}(N+M)$ down to ${\rm SU}(N) \times{\rm SU}(M)$. The states can then be organised into two types of multiplets; a massless vector multiplet transforming in the adjoint of the residual gauge group, and a massive vector multiplet in the bifundamental of ${\rm SU}(N)\times {\rm SU}(M)$. Massive multiplets are in the so-called $1/2$-BPS, ultrashort massive representations of $\mathcal{N}=4$ with central extension $Z_{IJ}=2M\Omega_{IJ}$, with Sp$(\mathcal{N}/2)$ $R$-symmetry, with skew form $\Omega_{IJ}$ and indices $I,J=1,\dots,\cN=4$.
For massless multiplets on the other hand, $R$-symmetry is enhanced to a full ${\rm SU}(4)$ by the vanishing of the central extension as $m\rightarrow0$.
Employing the notation of the previous section, we can combine the supercharges into a Dirac spinor $Q_I^A=\big(Q_{ \alpha I}, Q^{\dagger \dot{\alpha}}_{I}\big)$, such that the supersymmetry algebra takes the compact form
\begin{gather*}\label{eq:susy-algebra}
\{Q_{A I}, Q_{B J}\}=2\Omega_{I J}D^m_{AB}.
\end{gather*}
In the massless case, the structure of the supersymmetry algebra greatly simplifies as the only non vanishing component of the Dirac operator is $D_{\alpha\dot\alpha}^0=\nabla_{\alpha\dot\alpha}$.

The action of the supercharges arranges the states in multiplets as follows. The massive multiplet is composed of a massive spin one field $F_{AB}$, five massive scalars $\phi_{IJ}$ and four massive Weyl--Majorana spinors $\Psi_A^I$:
\begin{gather}\label{N=4massive}
\mathscr{F}^m=\big(\phi_{IJ}=\phi_{[IJ]},\Psi_{I}^ A,F^{AB}=F^{(AB)}\big), \qquad \phi_{IJ}\Omega^{IJ}=0.
\end{gather}
The massless multiplet is
\begin{gather}\label{N=4massless}
\mathscr{F}^0=\big(\phi_{IJ}=\phi_{[IJ]},\Psi^I_\alpha, \tilde\Psi_{I\dot \alpha},F_{\alpha\beta},F_{\dot\alpha\dot\beta}\big),
\end{gather}
where the $R$-symmetry indices now label the fundamental of ${\rm SU}(4)$ and can therefore no longer be raised and lowered.
It contains the two familiar ${\pm1}$ helicity states of the massless spin-1, six real massless scalars $\phi_{IJ}$ and eight massless gluino states via the chiral parts of $\Psi_\alpha^I$, $\tilde \Psi_{I \dot\alpha}$. We~note that the massless scalars $\phi_{IJ}$ are no longer trace-free; the extra 6th component arises from the loss of one of the polarization degrees of freedom going from the massive spin-1 field~$F_{AB}$ to the massless case.

For momentum eigenstates with space-time dependence $\phi=\exp({\rm i}k\cdot x)$, the supersymmetry generators reduce to the massive little group as
\begin{gather*}
Q_{AI}=\kappa_A^aQ_{aI}, \qquad \{Q_{aI},Q_{bJ}\}=2\Omega_{IJ}\varepsilon_{ab},
\end{gather*}
where $\kappa_a^A$ is defined by \eqref{plane-wave},
because the Dirac operator reduces as $D_{AB}^m\phi =(\kappa_A\kappa_B)\phi$. In~the massless limit we have the natural embedding of the little group via $\kappa_{1 \alpha}=0=\tilde{\kappa}_{0 \dot{\alpha}}$, $\kappa_{0 \alpha}=\kappa_\alpha$, $\tilde{\kappa}_{1 \dot{\alpha}}=\tilde{\kappa}_{\dot{\alpha}}$.

Both the massive and massless multiplets are annihilated by half of the supercharges so that their $8$ bosonic and $8$ fermionic states can all be encoded into the exterior powers of $\cN=4$ fermionic supermomenta $q_I$, $I=1,\dots,4$. These are defined to be the eigenvalues of an anticommuting subset of the $Q_{Ia}$. To define this subset, we introduce a basis $(\epsilon_a,\xi_a)$ of the fundamental representation of ${\rm SL}(2)$ so that the supermomenta are defined by the action of the supercharges on functions on on-shell superspace via
\begin{gather}\label{super-mom}
Q_{a I}\tilde{\mathscr{F}}(\kappa,q)=\left(\xi_{a} q_{I}+\epsilon_{a} \Omega_{I J} \frac{\partial}{\partial q_{J}}\right)\tilde{\mathscr{F}}(\kappa,q).
\end{gather}

The massive and massless multiplets are expanded on on-shell superspace as follows
\begin{gather}
\tilde{\mathscr{F}}_{(\kappa, q)}^{(m)} =F^{\epsilon \epsilon}(\kappa)+q_{I} \Psi^{\epsilon I}(\kappa)+q^{2} F^{\epsilon \xi}(\kappa)+\frac{1}{2} q_{I} q_{J} \Phi^{I J}(\kappa)+q^{2} q_{I} \Psi^{\xi I}(\kappa)+q^{4} F^{\xi \xi}(\kappa) ,\nonumber
\\
\tilde{\mathscr{F}}_{(\kappa, q)}^{(0)}=g^h(\kappa)+q_{I} \Psi^{\epsilon I}(\kappa)+\frac{1}{2} q_{I} q_{J} \varphi^{I J}(\kappa)+q^{2} q_{I} \Psi^{\xi I}(\kappa)+q^{4} g^{-h}(\kappa),
\label{CB-massive-massless-multiplet}
\end{gather}
with $q^4= (q_Iq_J)\big(q^Iq^J\big)$ and $\big(q^3\big)^I_a=\p q^4/\p q^a_I$.

It is then standard procedure to encode such multiplets in superfields on a supersymmetric extension of Minkowski space satisfying~\eqref{super-mom} and to derive a supersymmetric Penrose transform by establishing a supergeometric correspondence with super-twistor space. We can bypass some of this by studying the action of supersymmetry on super-twistors.

{\bf Supertwistors and the Penrose transform.}
We extend the bosonic complexified twistor $Y\in\T^\C$ with $\cN$ fermionic coordinates $\eta^I$, $I=1,\dots , \cN$ to give $\fZ=\big(\lambda_A,\mu^A, \eta^I\big)\in \cT^\C$, using Dirac-spinor notation.
These fermionic coordinates allow the supersymmetry to act geometrically~as
\begin{gather*}
Q_{AI}= \lambda_A \frac{\p}{\p\eta^I}+\eta^J\Omega_{JI}\frac{\p}{\p \mu^{A}}, \qquad \{Q_{AI}, Q_{BJ}\}=2\Omega_{IJ}\lambda_{[A}\frac{\p}{\p\mu^{B]}},
\end{gather*}
where the anticommutator now generates the action of translations on $\cT^\C$. This extends in the obvious way to the two-twistor description of supersymmetric massive particles in terms of~$\fZ_a$ with sums over the $a$-index in each term. Again, the supersymmetric extension for the $\fZ_a$ becomes:
\begin{gather*}
\mathcal{Y}_a=\big(\lambda_{aA}, \mu^A_a, \eta^I_a\big),
\end{gather*}
with again $I=1,\dots,\cN$. The plane wave representative for particles with spinor helicity data $\kappa_{Aa}=\big(\kappa_{\alpha a},\tilde \kappa^{\dot \alpha}_{ a}\big)$, supermomentum $q$ and polarization data $\epsilon_a$ will take the form:
\begin{gather}\label{Susy-plane-wave}
\Phi_{(\kappa,q)}(\mu,\lambda)=\int {\rm d}^2u\,{\rm d}^2v w \bar\delta^4((u\lambda_A)-(v\kappa_A)) \bar\delta((\epsilon v)-1){\rm e}^{{\rm i}u_{a}(\mu^{Aa} \epsilon_{A} +q_{I}\eta^{Ia})-\frac 1 2 (\xi v) q^2}.
\end{gather}
Here $w$ is a function of weight $2$ in $\fZ_a$(or $-2$ in $u$); as far as the Penrose transform is concerned, this can be taken to be $w= \big(\lambda_\alpha\tilde\lambda_{\dot\alpha}\big) {\rm e}^{\alpha\dot\alpha}$, where ${\rm e}^{\alpha\dot\alpha}$ is a polarization vector for the spin-1 1-form. In the string model however, $w$ plays an important role in the vertex operators, and will be constructed differently. This representative indeed satisfies
\begin{gather*}
 Q_{AI} \Phi_{\kappa,q}(\fZ_a) =\bigg((\kappa_A\xi)q_I+\epsilon_A\Omega_{IJ}\frac{\partial}{\partial q_I}\bigg)\Phi_{\kappa,q}(\fZ_a),
\end{gather*}
and we can then read off the Penrose transform for the component fields from the action of the supersymmetry generators.

\section{Massive two-twistor string}\vspace{1ex}
The significance of the twistor representations of spaces of massless and massive particles is that they are represented as symplectic quotients of vector spaces. This means that in order to construct a theory of maps from a Riemann surface $\Sigma \rightarrow \cP^\C_m$, we can start with a quantum field theory of maps $\Sigma \rightarrow\T^\C$ in the massless case and $\Sigma \rightarrow \T^\C\times \T^\C$ in the massive case; in both cases, by virtue of the twistor representations, these are free field theories on the worldsheet $\Sigma$. We~then realize the symplectic quotient in the Lagrangian framework by gauging the Hamiltonian symmetries as we shall describe below. These gauge symmetries are then dealt with via BRST in the quantum field theory.
In both cases, the free field theory action is based on the restriction of the symplectic potential $\Theta$ to $T^{0,1}_\Sigma$. This has the consequence that the worldsheet commutators and OPEs encode the symplectic structure $\Omega_m$ on $\mathcal{P}_m^\C$.

We first briefly review the massless case; although the construction for the massive two-twistor string will be analogous, but with target $\mathcal{P}_m^\C$ and the different massive supersymmetry representation.
 In the next section we explain how the models allow us to construct amplitudes as
correlation functions of vertex operators in these models.

{\bf The massless case.}
The twistor strings of Witten~\cite{Witten:2003nn} and Berkovits \cite{Berkovits:2004hg}
and the 4d ambitwistor string of \cite{Geyer_2014} are theories of holomorphic maps $\fZ=\big(\cZ,\tilde \cZ\big)\colon\Sigma \rightarrow \T^\C$
 gauged by $\C^*$, where we now use supertwistors $\cZ=\big(\lambda_\alpha,\mu^{\dot\alpha}, \eta^I\big)\in\T=\C^{4|\cN}$, $I=1,\dots, \cN$ and their complexification $\T^\C=\T\times \T^*$; this is the complexification of the four-dimensional massless Ferber superparticle~\cite{Ferber:1977qx}.\looseness=1

The four-dimensional ambitwistor string of \cite{Geyer_2014} is closest to the massive case, being with worldsheet fields twisted to take values in $\Omega^{\nicefrac12,0}_\Sigma$ and so we briefly review it here. It is a theory of holomorphic maps from a Riemann surface $\fZ\colon\Sigma\rightarrow \T_\C\otimes \Omega^{\nicefrac{1}{2},0}_\Sigma$, so that the coordinates $\fZ$ are worldsheet spinors. The reduction to ambitwistor space is enforced by gauging the little-group Hamiltonian $\fZ\cdot \fZ:=\cZ\cdot \tilde \cZ$ with the worldsheet gauge field $A\in \Omega^{0,1}_\Sigma$.
The basic bosonic 4d ambitwistor action in conformal gauge\footnote{The full action would start with a term $e T$, where $e\in T^{1,0}_\Sigma\otimes \Omega^{0,1}_\Sigma$ is a Beltrami differential thought of as a gauge field parametrizing complex structures on $\Sigma$ up to coordinate transformations and $T\in \big(\Omega^{1,0}_\Sigma\big)^2$ is the holomorphic stress energy tensor; this is then gauge fixed, giving rise to ghosts $(b,c)\in \big(\big(\Omega^{1,0}_\Sigma\big)^2,T_\Sigma^{1,0}\big)$ and BRST operator $Q=\oint c T + b c \p c/2$. } is based on the symplectic potental
\begin{gather}\label{massless-model}
S^0_{4d}=\int_\Sigma \cW\cdot \dbar \cZ -\cZ\cdot \dbar \cW+A\,\fZ\cdot \fZ .
\end{gather}
Classically, $A$ is a Lagrange multiplier that enforces the constraint $\fZ\cdot\fZ=0$ and the quotient by its Hamiltonian vector field arises because $\big(Z,\tilde Z\big)\rightarrow \big(\alpha Z,\alpha^{-1} \tilde Z\big)$ are gauge symmetries of the action when accompanied by the gauge transformations $A\rightarrow A+ \dbar \log \alpha$. Thus the holomorphic symplectic quotient to~$\A$ in~\eqref{A-def} is realized in this Lagrangian framework by the gauge field~$A$. In~the QFT this is implemented via BRST quantization. The models of~\cite{Geyer_2014} also include additional worldsheet matter fields but these are much as described for the massive case below.

{\bf Massive models.} In order to have target space $\cP_m^\C$, we start with maps $\fZ_a\colon \Sigma \rightarrow \T_\C\times \T_\C$, with the reduction to $\cP^\C_m$ obtained by gauging the complexified two-twistor massive internal symmetry group.
Thus, our theory is one of maps $\fZ_a\colon\Sigma \rightarrow \T_\C\otimes \C^2\otimes \Omega^{\nicefrac{1}{2},0}_\Sigma$ with action (again in conformal gauge)
\begin{gather*}
S_{4d}=\int_\Sigma \fZ^a\dbar \fZ_a+ A_{ab}\fZ^{a}\cdot \fZ^b+A \big(\lambda^2-j^H\big)+ \tilde A \big(\tilde{\lambda}^2-j^H\big) + S_m.
\end{gather*}
Here $a=1,2$ is the little group index, and $\big( A_{ab}=A_{(ab)},A, \tilde A\big)$ are worldsheet $(0,1)$-forms that act as Lagrange multipliers for the constraints, and as gauge fields for the internal two-twistor symmetry group. With this symmetry, we no longer have the freedom to allow worldsheet fields of different degrees as we did for the twistor-string.
In order to describe specific space-time theories, the basic action must be supplemented by further worldsheet fields such as a current algebra for gauge theory and some analogue of worldsheet supergravity for gravity with details given below.
Here we assume that it contains a current-algebra that gives rise to a $(1,0)$-form~$j_H$ on the worldsheet that generates some symmetry.

 To be more explicit, in quantizing the fields $\fZ_a(\sigma)=\big(\cZ_a(\sigma),\tilde \cZ_a(\sigma)\big)$, for $\sigma$ a coordinate on~$\Sigma$, the only non-trivial OPEs are
 \begin{gather*}\label{Y-OPE}
 \cZ^\cA_a(\sigma)\tilde \cZ_{b\cB}(0)=\frac{\delta^\cA_\cB}{\sigma}\varepsilon_{ab}+\cdots
 \end{gather*}
reflecting the Poisson brackets.
 These OPEs can lead to anomalies for the little group ${\rm SL}(2,\C)$ generated by $J^{ab}=\fZ^{a}\cdot \fZ^b$. For a consistent model these anomalies have to vanish, which requires judicious choices for the worldsheet matter $S_m$.

The fields $a$, $\tilde a$ gauge the constraints $\lambda^2-j^H=0=\tilde{\lambda}^2 - j^H$. These equations constrain the mass operators
\begin{gather*}
\lambda^2:
=\frac{1}{2}\lambda^a_\alpha\lambda_a^\alpha=\det(\lambda_\alpha^a),\qquad
\tilde{\lambda}^2:=\frac{1}{2}\tilde\lambda_{\dot\alpha}^a\tilde{\lambda}_a^{\dot\alpha} =\det\big(\tilde{\lambda}_a^{\dot\alpha}\big),
\end{gather*}
to be given by a $(1,0)$-form $j^H$ on the worldsheet $\Sigma$. We write $j^H$ to indicate that this will be taken to be the current associated to the element $h\in\mathfrak{g}$, living in the Cartan subalgebra of some symmetry of the system. This~$j^H$ will be constructed from the matter fields and, through the constraints above, will determine the masses of the particles. For a given matter content, different choices of~$j^H$ correspond to different distributions of masses within the models.
The massless models~\eqref{massless-model} are recovered from these massive ones when $j^H=0$ by reducing the path integral.

{\bf Worldsheet matter.}
A variety of physically interesting models can be constructed from different choices of~$S_m$. These will be made up of current algebras, whose action will be denoted by~$S_C$, and worldsheet fermions providing a supersymmetric extension of the worldsheet gauge algebra, denoted by~$S_\rho$. The latter will play a similar role to worldsheet supergravity in the superstring, and is requried for models describing gauge theory and supergravity.

A worldsheet current algebra is a theory on the worldsheet from which one can construct worldsheet currents $j^a\in \Omega^{1,0}_\Sigma\otimes \mathfrak{g}$ for some Lie algebra $\mathfrak{g}$, satisfying the OPE
\begin{gather*}
j^\mathfrak{a}(\sigma) j^\mathfrak{b}(0)\sim \frac{l\, \delta^{\mathfrak{ab}}}{\sigma^2} + \frac{f^{\mathfrak{ab}}_\mathfrak{c} j^\mathfrak{c}}{\sigma},
\end{gather*}
where $\mathfrak{a}$, $\mathfrak{b}$ are Lie-algebra indices, $l\in \Z$ is the level and $f^{\mathfrak{ab}}_\mathfrak{c}$ the structure constants of $\mathfrak{g}$.
Such current algebras can be constructed in a number of ways, most easily for SO$(n)$ and ${\rm SU}(n)$ by ``real'' or ``complex'' free fermions on the worldsheet. See also \cite{Geyer:2015jch} for a construction referred to as a comb-system, with level zero and novel properties that allow the construction of Einstein-Yang--Mills amplitudes. We will not specify the action $S_C$ explicitly, but merely assume that we have the currents $j^\mathfrak{a}$ in the theory.

For gauge and gravity theories, we need a supersymmetric extension of the worldsheet gauge algebra. This plays a similar role to the worldsheet supergravity of the conventional RNS models, see also \cite{Mason:2013sva} for the ambitwistor-string version. The supersymmetric extension of the bosonic gauge algebra $\mathfrak{sl}_2\times \mathbb{C}^2$ is constructed by introducing the worldsheet fermions $\big(\rho_A,\tilde\rho^A\big) \in
\Omega^0\big(\Sigma, K_\Sigma^{1/2}\big)$
with action
\begin{gather}
S_{\rho}=\int_\Sigma \tilde \rho^\sA\dbar \rho_\sA + b_a\big(\gamma_5^{\sA\sB}\lambda_\sA^a \rho_\sB\big)+\tilde b_a \lambda_{\sA}^a \tilde\rho^\sA.
\end{gather}
Here the $\big(b^a,\tilde b^a\big)$ are fermionic gauge fields and so are $(0,1)$-forms on the worldsheet. They are Lagrange multipliers that impose the constraints $\gamma_5^{\sA\sB}\lambda_\sA^a \rho_\sB = \lambda_{\sA}^a \tilde\rho^\sA = 0$ and their gauge transformations translate $\mu^A_a$ in the direction of $\big(\rho^A,\tilde \rho^A\big)$. The only non-trivial OPE's of the constraints are given by
\begin{gather}\label{WS-Susy}
 \big(\gamma_5^{\sA\sB}\lambda_\sA^a \rho_\sB\big)(z)\; \big(\lambda_\sB^b \tilde\rho^\sB\big)(w)\sim\frac{\varepsilon_{ab}}{z-w}\big(\lambda^2 -\tilde\lambda^2\big).
\end{gather}
These symmetries thus give a supersymmetric extension of the two-twistor internal symmetry group $\mathbb{C}\times \mathfrak{sl}_2\ltimes H(0,4)$, where $H$ denotes the Heisenberg Lie superalgebra.\footnote{The Heisenberg superalgebra $H(m_b,m_f)$ has a central element $z$, as well as $2m_b$ even and $m_f$ odd generators, $H=\la x_1,\dots, x_{2m_b},z \ra\oplus \la \psi_1,\dots, \psi_{m_f}\ra$. The generators satisfy the ``usual'' commutation relations
\[
 [x_i,x_{2i}]=z,\qquad \{\psi_r,\psi_s\}=2\delta_{rs} z.
\]
}

{\bf Models.} With these ingredients, models without ${\rm SL}(2,\C)$-anomalies can be constructed by combining a pair of worldsheet matter systems, much along the lines of the double copy for the RNS ambitwistor strings as in \cite{Geyer:2015jch} as follows:
\begin{alignat*}{3}
 & \text{massive bi-adjoint scalar}
\qquad && S^{\scalebox{0.7}{BAS}} = S_{4d}+S_C + S_{\tilde C},&\\
 &\text{super Yang--Mills on the Coulomb branch}
 \qquad && S^{\scalebox{0.7}{CB}} \hspace{4pt}= S_{4d}+S_\rho+S_C,& \\
 &\text{super-gravity}
 \qquad && S^{\scalebox{0.7}{sugra}} = S_{4d}+S_{\rho_1}+ S_{\rho_2}.&
\end{alignat*}

\noindent
In this construction two points are worth highlighting:{\samepage
\begin{itemize}\itemsep=0pt
\item[$(i)$] The closure of the constraint algebra requires that both constraints $\lambda^2-j^H=0=\tilde{\lambda}^2 - j^H$ involve the \emph{same} current $j^H$ for super Yang--Mills, whereas a more general construction is possible for the bi-adjoint scalar.

\item[$(ii)$] Unlike the twistor- and ambitwistor models for 4d massless theories, these models fit neatly into the double copy format \cite{Bern:2010ue} expressed directly in the CHY formulae \cite{Cachazo:2013iaa} and in the corresponding RNS ambitwistor strings \cite{Mason:2013sva}.
However, it is harder to find a $j^H$ to endow our particles with mass in the gravitational case because there is no additional current algebra, and with $j^H=0$ our models are massless. We also note that as in \cite{Berkovits:2018jvm, Geyer:2020iwz, Mason:2013sva}, both $S^{\scalebox{0.7}{CB}}$ and $S^{\scalebox{0.7}{BAS}}$ also contain a gravity sector, but it is of higher order and remains massless.
\end{itemize}}

{\bf BRST and anomalies.}
Gauge fixing the action via BRST generates ghost systems, the well-known $(b,c)\in \big(\Omega^1_\Sigma\big)^2 \times T_\Sigma $ for worldsheet diffeomorphisms, as well as additional fermionic ghosts associated to internal two-twistor symmetry group, and bosonic ghosts for the fermionic currents in $S_\rho$.
The BRST operator takes the usual form:
\begin{gather}\label{Q-BRST}
 Q=\oint c^{i}\bigg(T^m_{i}+\frac{1}{2}T^g_{i}\bigg),
\end{gather}
where the sum runs over all sets of ghosts, and $T^m$ and $T^g$ are the matter and ghost parts of the currents respectively.
By construction $Q^2=0$ classically, but in the QFT double contractions (or worldsheet bubble diagrams with two external gauge fields) can lead to anomalies so that $Q^2\neq 0$ with a potential obstruction arising from any of the gauged symmetries.
Here we briefly summarize the results of such calculations.

The models above only have a vanishing ${\rm SL}(2,\mathbb{C})$ anomaly (corresponding to the two-twistor internal symmetry group) for maximal space-time supersymmetry, as evident from the anomaly coefficient
\begin{gather*}
 \mathfrak{a}_{\scalebox{0.7}{$\mathrm{SL}(2)$}}=\sum_{i}(-1)^{F_i}\tr_{R_i}\big(t^kt^k\big) =
 \begin{cases}
 4\, \tr_{\scalebox{0.7}{F}}\big(t^kt^k\big) - \tr_{\scalebox{0.7}{adj}}\big(t^kt^k\big) = 0 & \text{bi-adjoint scalar},
 \\
 \frac{3}{4}(4-\mathcal{N})& \text{Coulomb branch},
 \\
 \frac{3}{4}(8-\mathcal{N})& \text{supergravity}.
 \end{cases}
\end{gather*}
The anomaly coefficient vanishes trivially for the bi-adjoint scalar, and for maximal supersymmetry in the case of gauge theory and gravity. The sum here runs over all fields that transform non-trivially under the internal two-twistor symmetry group $\mathrm{SL}(2,\mathbb{C})$. Similarly, the Virasoro central charge can be calculated for all models, giving
\begin{gather*}
\mathfrak{c}^{\scalebox{0.6}{BAS}} = -40 + \mathfrak{c}_j,
\qquad
 \mathfrak{c}^{\scalebox{0.6}{CB}}=-32 +\mathcal{N}+\mathfrak{c}_j,
\qquad
 \mathfrak{c}^{\scalebox{0.6}{sugra}}=-20 +\mathcal{N},
\end{gather*}
where $\mathfrak{c}_j$ denotes the central charge of the internal current algebra. The conformal anomaly thus vanishes for suitable choice of $S_j$, with $\mathcal{N}=4$ and $\mathfrak{c}_j=28$ for Yang--Mills theory on the Coulomb branch, and $\mathfrak{c}_j=40$ for the bi-adjoint scalar. The Virsaoro anomaly for the supergravity model also vanishes if we include a central charge term $\mathfrak{c}_{6d}=12$ arising from six compactified dimensions.
After BRST gauge-fixing, all such models are free worldsheet theories with vanishing anomalies.\footnote{Strictly speaking, the Lie algebra element $H$ should also be null or the current algebra should have level $l=0$ as in the comb systems of \cite{Casali:2015vta}.} We now explain how to obtain $n$-point amplitudes from these models.

\section{Amplitudes and vertex operators}
\label{sec:amplitudes}
In string theory, amplitudes are constructed as correlation functions of \emph{vertex operators}, one for each external particle. Vertex operators come in various forms that depend on how residual gauge freedom is fixed after initial gauge fixing, see for example \cite{Witten:2012bh}. The easiest to understand are the \emph{integrated vertex operators} that require integration over $\Sigma$; these are the generic case and arise as the perturbations of the action corresponding to infinitesimal background plane-wave fields. However, one also needs a small number of \emph{fixed vertex operators} that fix residual worldsheet diffeomorphisms and other symmetries such as those associated to the $a$, $\tilde{a}$ and $b$,~$\tilde{b}$ fields. We here just give brief details of the basic recipe required for the amplitude formulae and refer to further work \cite{wip} for full details of the derivation.

\subsection{Vertex operators}
Just as for the original twistor string, vertex operators for the two-twistor string are constructed from the Penrose-transform of plane waves $\Phi_\kappa(Y_a)\in H^1\big(\cP^\C_m,\cO(-2)\big)$ \eqref{H1_plane-wave}, or its supersymmetric extension $ \Phi_{\kappa,q}(\fZ_a)$~\eqref{Susy-plane-wave} with an appropriate choice of $w$. The identification of the $Y_a$ or $\fZ_a$ as spinors on the worldsheet implies that $\Phi_\kappa(Y_a)$ restricts to~$\Sigma$ to define an element of $H^{0,1}(\Sigma,T_\Sigma)$. Similar considerations apply to $ \Phi_{\kappa,q}(\fZ_a)$ but the line bundle on $\Sigma$
depends on the choice of~$w$. We will therefore choose~$w$ so as to take values in $\big(\Omega^{1,0}\big)^2$ so that $w\circ \Phi$ defines a~$(1,1)$-form on $\Sigma$ and can be integrated; here the notation $\circ$ indicates schematically that $w$ may depend on $u$, and should be included under the integral in $\Phi_\kappa$.
Thus fixed and integrated vertex operator can be constructed by including a state- and theory dependent factor $w\in\Omega^0\big(\Sigma,K_{\Sigma}^2\big)$ respectively as
\begin{gather}\label{VO-w-wave}
 V=c \; w\circ\Phi_\kappa(\sigma) ,\qquad
\mathcal{V}=\int {\rm d}\sigma \, w \circ \Phi_\kappa(\sigma) .
\end{gather}
Here $c$ is the fermionic ghost with values in $T^{1,0}\Sigma$ associated to worldsheet diffeomorphisms.

Vertex operators are required to be BRST $Q$-closed which implies invariance under the symmetries. For example, the presence of the $c$-ghost in $V$ gives $Q$-closure trivially for that part of~$Q$ containing $c$ which is associated to worldsheet diffeomorphisms, whereas, on the other hand, the integration of the $(1,1)$-form on $\Sigma$ that defines $\cV$ is manifestly diffeomorphism invariant. We~also expect to see similar invariance of both fixed and integrated vertex operators under the gauged two-twistor internal symmetry and the gauged fermionic symmetries in $S_\rho$ if present.
BRST-closed vertex operators can thus be constructed by including ghost factors for the remaining generators, including fermionic ghosts $t$, $\tilde t$ associated to the mass constraints gauged by~$A$,~$\tilde A$, and bosonic ghosts $\gamma^a$, $\tilde\gamma^a$ corresponding to $b^a$, $\tilde b^a$ respectively.

It is this $Q$-invariance that will fix the masses of our external particles. In the sum over ghosts in \eqref{Q-BRST}, $Q$ contains the term for the ghosts $t$ and $\tilde t$ that arose from gauge fixing the~$A$,~$\tilde A$:
\begin{gather*}
Q_m:= \oint t \big(\lambda^2-j^H\big) +\tilde t \big(\tilde \lambda^2 -j_H\big).
\end{gather*}
When acting on a vertex operator built as in~\eqref{VO-w-wave}, the operator $\lambda^2$ becomes the mass operator, yielding the $Q$-invariance condition that requires support on $\kappa^2=M^H$, where $M^H$ is the eigenvalue of the $H$-action on the vertex operator under~$j^H$.
Thus $Q$-invariance requires that the signed mass is given by the action of~$j^H$. This action is theory-dependent as follows.

{\bf Biadjoint scalars.} This theory involves two Lie algebras $\mathfrak{g}$, $\tilde{ \mathfrak{g}}$ with fields $\phi^{\mathfrak{a}\tilde{\mathfrak{a}}}$, where $\mathfrak{a}$, $\tilde{\mathfrak{a}}$ are the respective Lie-algebra indices. The action is $\int {\rm d}^4x\big( (\p\phi)^2 + f_{\mathfrak{a}\mathfrak{b}\mathfrak{c}}\tilde f_{\tilde {\mathfrak{a}}\tilde {\mathfrak{b}}\tilde {\mathfrak{c}}}\phi^{\mathfrak{a}\tilde{\mathfrak{a}}}\phi^{\mathfrak{b}\tilde{\mathfrak{b}}} \phi^{\mathfrak{c}\tilde c}\big)$, where the~$ f_{\mathfrak{a}\mathfrak{b}\mathfrak{c}}$,~$\tilde f_{\tilde {\mathfrak{a}}\tilde {\mathfrak{b}}\tilde {\mathfrak{c}}}$ are the structure constants of the respective Lie algebras.
We will also incorporate a~mass term $ \tr([\phi,H])^2$, where $H\in \mathfrak{g}\oplus \tilde{ \mathfrak{g}}$.
Each particle then has a pair of ``colours'' $\big(\mathfrak{t}_\mathfrak{a},\tilde{ \mathfrak{t}}_{\tilde{\mathfrak{a}}}\big)\in \mathfrak{g}\oplus \tilde{\mathfrak{g}}$ and momentum $k$ with mass determined by the eigenvalues of the action of $H$ in the form of the plane wave $\phi^{\mathfrak{a}\tilde{\mathfrak{a}}}=\mathfrak{t}^{\mathfrak{a}}\tilde{\mathfrak{t}}^{\tilde {\mathfrak{a}}}\e^{{\rm i}k\cdot x}$. These theories don't admit supersymmetry so we will simply use~\eqref{H1_plane-wave} and the two colours $\mathfrak{t}^{\mathfrak{a}}\tilde{\mathfrak{t}}^{\tilde {\mathfrak{a}}}$ to build the corresponding vertex operators, so that the mass is assigned via
\begin{gather}\label{group-action-on-vo}
 j^H(z)\; \mathfrak{t}_{\mathfrak{a}} j^{\mathfrak{a}}(w)\sim \frac{M^H_\mathfrak{t}}{z-w}\mathfrak{t}_{\mathfrak{a}} j^{\mathfrak{a}}.
\end{gather}

{\bf Coulomb branch of maximal super-Yang--Mills.} Here one takes the standard $\cN=4$ super-Yang--Mills with Lie algebra $\mathfrak{g}$, and assumes that the scalars of the massless sector $\phi_{IJ}$ are expanded around the constant $H\Omega_{IJ}$, where $H\in \mathfrak{g}$ is taken to be semi-simple. Decomposing $\mathfrak{g}$ into eigenspaces under the adjoint action of $H$, those with eigenvalue zero give rise to massless multiplets \eqref{N=4massless}, but those with non-zero eigenvalue $M^H$ acquire a mass $|M^H|$ and become massive multiplets as in \eqref{N=4massive}, where the scalar that is a multiple of $\Omega_{IJ}$ becomes a polarization state of the massive spin-1 field via the Higgs mechanism.

To be more explicit, consider the symmetry breaking $U(N+M)\rightarrow U(N)\times U(M)$, obtained by choosing
$H \sim \mathrm{diag}(\mathds{1}_{N},0_M)$. This leaves two types of states in the reduced theories, the first corresponding to $\mathfrak{t}\in \mathfrak{u}_N\times \mathfrak{u}_M$, the diagonal blocks that commute with $H$, and the second the off-diagonal blocks consisting of $\mathfrak{m} \in \C^N\otimes \big(\C^M\big)^*\oplus \C^M\otimes \big(\C^N\big)^*$ for which
 $[H,\mathfrak{m}]= M^H_\mathfrak{m} \mathfrak{m}$.
The first type then corresponds to massless states, and the second will have mass $\big|M^H_\mathfrak{m}\big|$, following from the $Q$-invariance of the vertex operators as above.
As a consequence, the OPEs \eqref{group-action-on-vo} of the respective currents take the following form:
\begin{gather}\label{group-action-CB}
j^H(z)\; \mathfrak{t}\cdot j(w)\sim 0,\qquad
j^H(z)\;\mathfrak{m}\cdot j(w)\sim\frac{M^H_\mathfrak{m}}{z-w}\mathfrak{m}\cdot j.
\end{gather}
We can thus identify vertex operators built from currents $\mathfrak{t}\cdot j$ with the massless vector multiplet transforming in the adjoint of the residual $U(N)\times U(M)$ gauge group, whereas vertex operators built from $\mathfrak{m} \cdot j$ describe the massive vector multiplet.\footnote{The massive states are $1/2$-BPS, ultrashort massive representations of $\mathcal{N}=4$ with central extension \mbox{$|Z_{IJ}|=2m$}.}

{\bf Supergravity.} We are not able to introduce massive spin-2 fields with this mechanism. However, we can introduce masses elsewhere into the supermultiplet by choosing for $j^H$ a generator of the $R$-symmetry. While we will not develop this theme in any detail, a short discussion is included in Section~\ref{sec:discussion}. The graviton states remain in a massless supermutliplet with $\cN=4$, formed as spin-2 part of the tensor product of two copies of the gauge mutiplet \eqref{N=4massless}, supplemented by an ultrashort matter multiplet, again with $\cN=4$.

Because both the $A$, $\tilde A $ fields and $b^a$, $\tilde b^a$ fields have moduli that need to be fixed when computing correlators, we need to adjust some vertex operators to fix this residual gauge freedom. Here we will suppress these details and give only the equivalent of integrated vertex operators for these symmetries, and refer to the analogous calculations in the 5d ambitwistor string \cite{Geyer:2020iwz} or \cite{wip} for details.

The main new ingredient here required for BRST-invariance of the vertex operators are delta functions fixing the value of $\lambda^2$ and $\tilde\lambda^2$ to the relevant masses $M_i^H$ arising from the eigenvalues of $j^H$. These arise from integrating out the moduli of the associated gauge fields in the presence of vertex operators.
Thus the generic integrated vertex operators for these theories can be summarized as
\begin{gather}\begin{split}
w_{\scalebox{0.6}{BAS}}(\sigma_i) ={}&\delta\big( \mathrm{Res}_{\sigma_i}
\big(\lambda^2 -j^H\big)\big) \delta\big( \mathrm{Res}_{\sigma_i} \big(\tilde\lambda^2
 -j^H\big)\big) \mathfrak{t}_{\mathfrak{a}} j^{\mathfrak{a}} \tilde{\mathfrak{t}}_{\mathfrak{a}} \tilde{j}^{\mathfrak{a}},
 \\
 w_{\scalebox{0.6}{CB}} (\sigma_i) ={}&\delta\big( \mathrm{Res}_{\sigma_i}
\big(\lambda^2 -j^H\big)\big) \delta\big(\mathrm{Res}_{\sigma_i} \big(\tilde\lambda^2 -j^H\big)\big)
 \big(e_i^{\sA\sB}(\lambda_\sB\lambda_\sA)+\epsilon_i^\sA\epsilon_{i\sB} \rho_\sA\tilde\rho^\sB\big) \mathfrak{t}_{\mathfrak{a}} j^{\mathfrak{a}},
 \\
 w_{\scalebox{0.6}{grav}}(\sigma_i) ={}&\delta\big(\mathrm{Res}_{\sigma_i}
 \big(\lambda^2 -j^H\big)\big) \delta\big( \mathrm{Res}_{\sigma_i} \big(\tilde\lambda^2
 -j^H\big)\big)\big(e_i^{\sA\sB}(\lambda_\sA\lambda_\sB)-\epsilon_i^\sA\epsilon_{i\sB} \rho_\sA\tilde\rho^\sB\big)
 \\
 &\times \big(\tilde e_i^{\sA\sB}(\lambda_\sA\lambda_\sB)-\tilde \epsilon_i^\sA\epsilon_{i\sB} \rho_\sA\tilde\rho^\sB\big),
 \end{split}\label{eq:int_VO}
\end{gather}
where $e_i$ and $\tilde e_i$ are polarisation vectors. When evaluating correlators including these vertex operators, we will see that the delta-functions enforce the mass-shell condition. One can see from~(\ref{group-action-on-vo})--(\ref{group-action-CB}) that taking the residue of $j^H$ acting on a $H$-eigenstate extracts the mass~$M^H_i$.
Note that in the gravitational case the $M^H_i$ will be zero for any multiplet containing the gravitational spin two field.

\subsection{Massive amplitudes as two-twistor string correlators}

Amplitudes are calculated as correlators in the ambitwistor string models. Due to ghost zero modes and residual symmetry after gauge fixing, a non-trivial correlator with $n$ vertex operators must contain three fixed vertex operators $V_1$, $V_2$, $V_3$, not necessarily of the same type, whose details we suppress:
\begin{gather*}
\mathcal{A}_{n}=\bigg\langle V_{1}V_{2}V_{3}\prod_{i=4}^{n} \mathcal{V}_{i} \bigg\rangle.
\end{gather*}
After gauge fixing, all the fields are free and these correlation functions can be straightforwardly evaluated explicitly for any $n$ to give formulae for scattering amplitudes.

Each vertex operator contains an exponential from the $\Phi_\kappa$ factor as in the classes \eqref{Susy-plane-wave} embedded in the definition of the vertex operators \eqref{VO-w-wave}. Directly computing their correlation functions would be awkward, but in the expression of the correlation function as a path integral, the exponentials from the vertex operators can be absorbed as source terms into the action for the $\fZ_a$-path integral. Since the correlator is independent of $\mu^\sA_a$, and the action is linear in $\mu^\sA_a$, the path integrals over these fields can be performed directly, localizing the correlator onto the classical equations of motion, but now with sources arising from the vertex operators
\begin{gather*}
\bar{\partial} \lambda_{A}^{a}=\sum_{i=1}^{n} u_{i}^{a} \epsilon_{i A} \,\bar{\delta}(\sigma-\sigma_{i}),\qquad
\bar{\partial} \eta_{I}^{a}=\frac{1}{2}\sum_{i=1}^{n} u_{i}^{a}\,q_{i I} \,\bar{\delta}(\sigma-\sigma_{i}).
\end{gather*}
On the Riemann sphere, these are solved uniquely by
\begin{gather}\label{sol-path-integral}
\lambda_{A}^{a}(\sigma)=\sum_{i=1}^{n} \frac{u_{i}^{a} \epsilon_{i A}}{\sigma-\sigma_{i}},
\qquad
\eta_{I}^{a}(\sigma)=\frac{1}{2} \sum_{i=1}^{n} \frac{u_{i}^{a}\,q_{i I} }{\sigma-\sigma_{i}}.
\end{gather}
Using this, the remainder of the correlator can be evaluated as follows:
\begin{itemize}\itemsep=0pt

 \item
 \emph{Polarized scattering equations and measure.} Evaluating the ghost path integrals and collecting the various measure factors as well as the delta-functions from $\Phi_\kappa$ gives the following measure:
 \begin{gather}\label{measure}
{\rm d}\mu_{n}^{\mathrm{pol}}:=\frac{\prod_{j} {\rm d}\sigma_{j}\, {\rm d}^{2} u_{j} \,{\rm d}^{2} v_{j}}{\operatorname{vol} \mathrm{SL}(2, \mathbb{C})_{\sigma} \times \mathrm{SL}(2, \mathbb{C})_{u}} \; \prod_{i=1}^{n} \bar{\delta}^{4}\big(( u_{i} \lambda_{A}(\sigma_{i}))-( v_{i} \kappa_{i A})\big).
\end{gather}
 A short counting reveals that the integrals are fully localized on the constraints enforced by the delta-functions, known as the \emph{polarised scattering equations} $\mathcal{E}_{i\sA}=0$, where
 \begin{gather}\label{eq:pol-SE}
\mathcal{E}_{i\sA}:=(u_i\lambda_{A}(\sigma_i))-(v_i\kappa_{iA})=\sum_{j\neq i}\frac{(u_iu_j)\epsilon_{jA}}{\sigma_i-\sigma_j}-(v_i\kappa_{iA}).
\end{gather}
These equations and corresponding measure are by now well studied in \cite{Albonico:2020mge,Geyer:2018xgb}; the polarized scattering equations imply a massive analogue of the original scattering equations for the~$\sigma_i$ described in Section~\ref{sec:discussion}, and the measure reduces to the CHY measure \cite{Cachazo:2013iea}.
After performing the integrals, the amplitudes still contain six residual delta-functions, whose form can be extracted from
\begin{gather}\label{eq:def_K}
 \epsilon_{i[A}\mathcal{E}_{iB]}=\sum_iK_{iAB}=0,\qquad \text{where}\quad K_{iAB}:=(\kappa_{iA}\kappa_{iB})=
\begin{pmatrix}
\kappa_i^2 \varepsilon_{\alpha \beta} & k_{i\alpha}{ }^{\dot{\beta}} \\
-k_{i\beta}^{\dot{\alpha}} & \tilde{\kappa}_i^2 \varepsilon^{\dot{\alpha} \dot{\beta}}
\end{pmatrix}\!.
\end{gather}

The residual delta-functions thus encode conservation of momentum $k_{\alpha\dot\alpha}$, as well as restrictions on $\kappa^2=\det(\kappa_\alpha^a)$ and $\tilde{\kappa}^2$:\, $\sum_i \kappa_i^2=0=\sum_i\tilde{\kappa}_i^2$. Notice that the closure of the algebra of constraints \eqref{WS-Susy} imposes that $\kappa_i^2=\tilde{\kappa}_i^2$ so that these two conditions are actually equivalent.
One can show that, because each $M_i^H$ is the charge under a symmetry on a compact worldsheet, the amplitude vanishes unless $\sum_iM_i^H=0$. Then this condition, together with conservation of $\kappa^2$ and the $n-1$ mass assigning delta functions, entails that also for the remaining particle $\kappa_1^2=M_1^H$.
 \item \emph{Mass fixing.} The action of the operators $\lambda^2$ and $\tilde \lambda^2$ reduces to the residues of their zero-modes \eqref{sol-path-integral} at $\sigma_i$. From the above formula, on the support of the delta functions in~\eqref{measure}, this can be seen to yield $\kappa^2_i$ and $\tilde \kappa_i^2$. Thus the mass parameters for each particle~$\kappa_i^2$ and~$\tilde \kappa_i^2$ are fixed to the appropriate value $M_i^H$ via the delta function in that vertex operator in~\eqref{eq:int_VO}:
\begin{gather*}
\prod_{i=2}^n\delta\big(\mathrm{Res}_{\sigma_i}\lambda^2-M^H_i\big) \delta\big(\mathrm{Res}_{\sigma_i}\tilde\lambda^2-M^H_i\big)
=\prod_{i=2}^n\delta\big(\kappa_i^2-M_i^H\big) \delta\big(\tilde\kappa_i^2- M_i^H\big),
\end{gather*}
now clearly enforcing the mass-shell condition for $n-1$ particles. That for the $1$st particle is omitted in one of the fixed vertex operators (to fix a residual degree of gauge freedom) but arises instead as a consequence of the residual delta functions that gives
$\sum_i \kappa_i^2=0=\sum_i\tilde{\kappa}_i^2$ as discussed above.
 \item \emph{Supersymmetry.}
 In the supersymmetric cases, the vertex operators contain exponential factors in the $\eta^a_I$s; these localize onto the solution to \eqref{sol-path-integral}, to yield:
\begin{gather*}
{\rm e}^{F_\mathcal{N}}
:=\exp\Bigg(\sum_{j<k}\frac{(u_ju_k) q_j\cdot q_k}{\sigma_j-\sigma_k}-\frac{1}{2}\sum_{j=1}^n(\xi_jv_j)q_j^2\Bigg).
\end{gather*}
This factor carries all the dependence on the supermomenta and can be expanded to yield the various component-amplitudes according to the supermomentum expansion of the supermultiplet \eqref{CB-massive-massless-multiplet}. The form of this factor moreover guarantees that the amplitudes are supersymmetrically invariant \cite{Albonico:2020mge, Geyer:2018xgb}.
\end{itemize}
Combining these factors leads to the following expression for massive amplitudes for super Yang--Mills on the Coulomb branch and massive bi-adjoint scalar theory;
\begin{gather}\label{final-formula}
\mathcal{A}_{n}=\int {\rm d}\mu_{n}^{\mathrm{pol}} \, \mathcal{I}_{n} {\rm e}^{F_\mathcal{N}}.
\end{gather}
Here, $\mathcal{I}_n$ -- the ``integrand'' of the amplitude expression -- contains all the remaining correlator factors. Since the current algebra and Heisenberg superalgebra do not interact, $\mathcal{I}_n$ can be split further into two ``half-integrand'' factors that can be calculated individually. The current algebra correlator at leading trace gives the omnipresent Parke--Taylor factor,
\begin{gather*}
 \langle \mathfrak{t}_{\mathfrak{a}_1} j^{\mathfrak{a}_1}(\sigma_1)\cdots \mathfrak{t}_{\mathfrak{a}_n} j^{\mathfrak{a}_n}(\sigma_n)\rangle
 =\sum_{\alpha\in S_n/D_n}\frac{\operatorname{tr}\left(\mathfrak{t}_{\alpha(1)} \cdots \mathfrak{t}_{\alpha(n)}\right)}{\sigma_{\alpha(1)\alpha(2)}\cdots \sigma_{\alpha(n)\alpha(1)}}=:\sum_{\alpha\in S_n/D_n}\operatorname{PT}(\alpha),
\end{gather*}
with the sum running over dihedrally inequivalent orderings. The evaluation of the $\rho\tilde\rho$ half-integrand is also standard (cf.~\cite{Geyer:2020iwz}), giving the reduced determinant,
\begin{gather*}
\det{}'H:=\frac{1}{(u_1u_2)}\det H^{[12]}_{[12]} ,
\end{gather*}
where, the $n\times n$ matrix $H$ is defined by
\begin{gather*}
H_{i j}=\frac{\epsilon_{i A} \epsilon_{j}^{A}}{\sigma_{i j}}, \qquad
H_{i i}=-e_i^{\sA\sB}(\lambda_\sA\lambda_\sB)(\sigma_{i}),
\end{gather*}
and the sub- and superscripts indicate that both the rows and the columns 1 and 2 have been removed. It can be shown that the reduced determinant $\det{}'H$ is invariant under permutation of \emph{all} particle labels; this can be seen directly \cite{Albonico:2020mge, Geyer:2018xgb}, but also follows from its correlator origin. We thus arrive at the following integrands;
\begin{gather}\label{integrands}
 \mathcal{I}_{n}^{\scalebox{0.6}{BAS}}=\operatorname{PT}(\alpha)\operatorname{PT}(\beta),\qquad
 \mathcal{I}_{n}^{\scalebox{0.6}{CB}}=\operatorname{PT}(\alpha)\det{}'H, \qquad
 \cI_n^{\rm Grav}= \det{}'H\det{}'\tilde H.
\end{gather}
This completes the specification of the massive amplitude formulae \eqref{final-formula}. For worked out examples at three and four points, we refer the reader to the computations in six dimensions presented in \cite{Albonico:2020mge}. They reproduce the formulae present in the literature for two or four particles in the massive supermultiplet and the remaining particles massless.

\section{Summary and discussion} \label{sec:discussion}

We have therefore seen that a chiral string whose target is the complexification of Penrose's two-twistor representation of the massive particle phase space yields theories of massive particles in four dimensions. The spectrum of these models includes massive particles, and correlators give amplitude formulae for super Yang--Mills on the Coulomb branch among other theories. These string models represent the confluence of two separate developments: the twistor-particle program of the 70's describing massive particles, and the more recent ambitwistor string models describing scattering amplitudes for massless particles. In the latter approach a chiral or holomorphic string whose target is the complexification of the space of massless particles yields amplitudes for theories of massless fields. Here we have seen that the logic extends naturally to massive particles. The significance of Penrose's twistor description is that it provides a canonical representation of the space as the symplectic quotient of a vector space modulo a Hamiltonian group action allowing the BRST quantization of a free quantum string. The twistorial description furthermore facilitates the incorporation of fermions and supersymmetry.

From the path integral of these models, we have arrived at the compact formulae \eqref{final-formula}, supplemented by \eqref{measure} and \eqref{integrands} supported on a massive version \eqref{eq:pol-SE} of the polarised scattering equations with manifest supersymmetry for appropriate gauge and gravity theories including massive particles. Like all twistor-string \cite{Cachazo:2012kg, Roiban:2004ka}, CHY~\cite{Cachazo:2013hca} and ambitwistor-string amplitude formulae, all the integrations are saturated against delta functions so that these are really residue formulae summing contributions from the $(n-3)!$ solutions to a massive extension of the scattering equations discussed further below.
 As shown in~\cite{Albonico:2020mge}, the extra data in the polarised extension is uniquely obtained by linear equations on the support of these scattering equations and the formulae are linear in the polarization data.

The models described in this paper, and related ones, can also be derived via a symmetry reduction of the higher-dimensional ambitwistor string models~\cite{wip}. This alternative interpretation of the two-twistor string highlights that these worldsheet models describe a~subset of massive models, where the particle masses are related to their (higher-dimensional) charges under a~symmetry. While this may appear restrictive, it includes many theories of immediate interest; in particular, all massive particles that we encounter in the standard model arise from the Higgs mechanism that can be obtained by symmetry reduction.
The upcoming paper \cite{wip} will also contain full details of the fixed vertex operators and picture changing operators that we omitted for brevity.
Mirroring the close relation of the models, the resulting amplitude formulae for massive particles are also closely related to those obtained previously by dimensional reduction from six and five dimensions \cite{Albonico:2020mge,Geyer:2018xgb,Geyer:2020iwz}. These can further be related to the formulae of \cite{Cachazo:2018hqa} by a change of gauge choice of an embedding inside a Lagrangian grassmannian \cite{Schwarz:2019aat}.
At low point orders, these expressions match the results obtained in \cite{Arkani-Hamed:2017jhn,Craig_2011,Herderschee_2019,Johansson:2019dnu,Lazopoulos:2021mna,Ochirov_2018} by BCFW recursion.

Contrary to the massless four-dimensional formulae of \cite{Geyer_2014}, in which the double copy properties are hidden in the measure, the expressions derived here present the standard structure with two half integrands that can be combined to form amplitudes for scalars, spin-1 and spin-2 particles as in the
CHY formulae and corresponding RNS models of \cite{Casali:2015vta, Mason:2013sva}.
 The embedding of the massless models in these novel massive ones give a new outlook on the corresponding massless formulae.

We conclude by discussing further features and connections to related formulae for massive amplitudes and future directions that we plan to address in~\cite{wip}.

{\bf Massive scattering equations.} The massive polarised scattering equations \eqref{eq:pol-SE} are closely related to the massive scattering equations in the CHY framework proposed by Dolan \& Goddard and by Naculich \cite{Dolan:2013isa, Naculich:2015zha}. We first note that as remarked after \eqref{xdotk} that the delta functions imply:
\begin{gather*}
 K^{AB}_iP_{AB}(\sigma_i):=k_i^{\alpha\dot\alpha} P_{\alpha\dot\alpha} -m_i \lambda^2(\sigma_i)-m_i \tilde \lambda^2(\sigma_i)=0.
\end{gather*}
After the path-integral we have that $\lambda_{Aa}$ is given in \eqref{sol-path-integral} and we can compute \cite{Albonico:2020mge, Geyer:2018xgb}
\begin{gather}\label{eq:def_P}
 P_{\sA\sB}(\sigma):=(\lambda_A\lambda_B)= \sum_{i}\frac{K_{i\sA\sB}}{\sigma-\sigma_i}\,{\rm d}\sigma,
\end{gather}
where $K_i$ is defined as in~\eqref{eq:def_K}. When the momenta $K_i$ are null, $P_{\sA\sB}$ plays a crucial role in the vector representation of the ambitwistor string \cite{Mason:2013sva}, where a similar path integral calculation to the one in Section~\ref{sec:amplitudes} localizes $P$ onto~\eqref{eq:def_P}. Naculich then showed that certain massive amplitudes localize on a massive extension of the CHY scattering equations,
\begin{align*}
 0=K_{i\sA\sB} P^{\sA\sB}(\sigma_i)=
 \sum_{j\neq i}\frac{k_i\cdot k_j-M_iM_j}{\sigma_i-\sigma_j}.
\end{align*}

The polarised scattering equations $\mathcal{E}_{i\sA}$ then clearly imply Naculich's massive scattering equations via $K_{i\sA\sB} P^{\sA\sB}(\sigma_i)=\det\big(\kappa_{i\sA}^a,\lambda^{b\sA}(\sigma_i)\big)=0$, where the last equality holds on the support of~$\mathcal{E}_{i\sA}$.
Vector ambitwistor models that directly yield these equations will be presented in~\cite{wip}.

{\bf Masses from $\boldsymbol R$-symmetry.} Novel gauge and gravitational models with massive degrees of freedom -- albeit massless gluons and gravitons -- can be formulated by choosing $j^H$ from the Cartan of the $R$-symmetry. In particular, we may take $j^H = m_1 \tilde \eta^1_a\eta_1^a+ m_2 \tilde \eta^2_a\eta_2^a$ in the two-twistor models $S^{\scalebox{0.7}{CB}}$ and $S^{\scalebox{0.7}{sugra}}$. For generic parameters $m_1$ and $m_2$, this results in non-supersymmetric models with massive scalars and fermions (and spin-1 in the supergravity model); no mass can be assigned to the spin-2 field since it transforms trivially under the $R$-symmetry. It is also possible to preserve a residual $\cN=2$ supersymmetry for gauge theory (and $\cN=4$ for supergravity) by constructing the current $j^H$ from only one Cartan generator with $m_2=0$.

{\bf Loops from a gluing operator.} This two-twistor string provides an alternative formulation of the massless ambitwistor string \cite{Geyer_2014}, but in a framework in which a massless field can be \emph{deformed} to go off-shell. This allows us to adapt the elegant method of deriving loop amplitudes in \cite{Roehrig:2017gbt} via a \emph{gluing operator} but now to theories with fermions and supersymmetry such as super Yang--Mills theory.
This construction arises from the vector model of the ambitwistor string, where it has been shown that loop integrands can be localized on a \emph{nodal sphere}
rather than the torus that more usually arises in string theory \cite{Geyer:2015bja, Geyer:2015jch}.
At the level of the worldsheet model, this nodal structure of loop correlators is realized in the worldsheet CFT as the \emph{gluing operator} $\Delta$, which encodes the propagator of the target-space field theory. One-loop amplitudes then have two equivalent descriptions; a string-inspired one as correlators on a torus, and an alternative representation as $g=0$ correlators in the presence of a gluing operator.

\subsection*{Acknowledgements}
We are grateful to Alexander Ochirov for discussions and comments.
GA is supported by the EPSRC under grant EP/R513295/1. YG gratefully acknowledges support from the CUniverse research promotion project ``Toward World-class Fundamental Physics'' of Chulalongkorn University (grant reference CUAASC). LJM is grateful to the STFC for support under grant ST/T000864/1. We would also like to thank the SIGMA team for their courage and dedication, working through difficult circumstances in Kyiv in the face of the Russian invasion and aggression.

\pdfbookmark[1]{References}{ref}
\LastPageEnding

\end{document}